\documentstyle[aps,epsfig]{revtex} 

\begin{document}
\draft
\title{Influence of tunneling on electron screening in low energy nuclear reactions in laboratories}
\author{Sachie Kimura, $^{a}$ \thanks{%
E-mail address: sachie@nucl.phys.tohoku.ac.jp} Noboru Takigawa, $^{a}$ 
\thanks{%
E-mail address: takigawa@nucl.phys.tohoku.ac.jp} Masanori Abe $^{b}$ \thanks{%
E-mail address: mabe@isenshu-u.ac.jp} and David M. Brink $^{c}$ \thanks{%
E-mail address: brink@thphys.ox.ac.uk}}
\address{$^a$Department of Physics, \\
Graduate School of Science, Tohoku University, Sendai 980-8578, Japan}
\address{$^b$Department of Basic Science, Ishinomaki Senshu University, \\
Ishinomaki 986-8580, Japan}
\address{$^c$Theoretical Physics, University of Oxford,1 Keble Road, Oxford OX1 3NP,
UK}
\date{\today}
\maketitle

\begin{abstract}
Using a semiclassical mean field theory, we show that the screening
potential exhibits a characteristic radial variation in the tunneling region
in sharp contrast to the assumption of the constant shift in all previous
works. Also, we show that the explicit treatment of the tunneling region
gives a larger screening energy than that in the conventional approach, 
which studies the time evolution only in the classical region and 
estimates the screening energy from the screening potential at the 
external classical turning point. 
This modification becomes important if the electronic state is not a
single adiabatic state at the external turning point either by pre-tunneling
transitions of the electronic state or by the symmetry of the system even if
there is no essential change with the electronic state in the tunneling
region.
\end{abstract}

\bigskip

Nuclear reaction rates at low energies play the key role in energy
generation in stars and the primordial and stellar nucleosynthesis. 
The bare reaction rates
are modified in stars by the screening effects of  free and bound electrons.
The knowledge of the bare nuclear reaction rates at low energies is
important not only for the understanding of various astrophysical nuclear
problems, but also for assessing the effects of host material in low energy
nuclear fusion reactions in matter. This is currently a subject of great
interest in nuclear physics. Rolfs and his colleagues have reported that the
experimental cross sections of the $^{3}$He(d,p)$^{4}$He and of D($^{3}$He,p)
$^{4}$He reactions with gas target show \ an increasing enhancement with
decreasing bombarding energy with respect to the values obtained by
extrapolating from the data at high energies 
\cite{FIGAssenbaumLangankeRolfs}. 
They also claimed that the enhancement is larger in the 
$^{3}$He(d,p)$^{4}$He reaction. 
Since then similar enhancement has been reported for many
systems with not only gas targets, but also with metal targets such as the 
$^{6}$Li(p,$\alpha $)$^{3}$He reaction.

These observations have motivated many theoretical as well as experimental
studies. Many of them attempted to attribute the enhancement of the reaction
rate to the screening effects by bound target electrons.
A simple approach is to assume that the
screening effects can be well represented by a constant, i.e. radially
independent, decrease of the barrier height in the tunneling region. This
decrease is named the screening energy. It is determined by making a fit to the
data. A puzzle is that the screening energy obtained by this procedure 
exceeds the value in the so called adiabatic limit, 
which is given by the 
difference of the binding energies in the united atom and in the target atom
and is theoretically thought to provide the maximum screening energy, for all
systems so far studied experimentally \cite{rolfs95}
(see ref.\cite{raiola01} for a recent modification).
For $^{7}$Li(p,$\alpha $)$\alpha $ reaction, in addition to the direct
measurement, an indirect measurement of the cross section using the Trojan
horse method has recently been made \cite{clauscatania}. The comparison
between the two methods indicates again that the screening energy in the
direct method exceeds the adiabatic limit by a large factor. 

One should, however, keep in mind that 
the stopping power is not well established,
especially for gas target, at such low energies \cite{gs-e,gs-t,bp,raiola01}. 
Also different values of the screening energy are obtained depending 
on the method of analysis \cite{barker,junker}.

In this paper, we discuss the properties of the screening potential in the
tunneling region. We examine, in particular, whether it can be represented
by a constant shift as has been postulated  in all previous studies. We also
examine the validity of the former dynamical approach 
in refs.\cite{ShoppaKooninLangankeSeki,shoppa96}, 
which solves the coupled equations for the
electronic and nuclear motions only in the classical region, and estimates
the screening energy by using the electronic wave function at the external
classical turning point. To that end, we describe the time evolution of the
electrons by a Schr\"{o}dinger equation and the relative motion between the
projectile and target nuclei by classical Newtonian equations. They are
coupled to each other through a variational principle leading to a mean field
theory. In that sense, our formalism is the same as that of Shoppa et al. 
\cite{ShoppaKooninLangankeSeki,shoppa96} 
for the classical region. However, we extend
the study to the tunneling region as well.

We denote the coordinate of the relative motion between the projectile and
target nuclei by ${\bf R}$ and that of the electrons by $\xi$, which
contains in general the coordinate of the center of mass of electrons
relative to the center of mass of the target and projectile nuclei, as well
as their intrinsic coordinates. Considering the head on collision, we assume
the following Hamiltonian for the total system, 
\begin{equation}  \label{eq:totham}
H(R,\xi)=-\frac{\hbar^2}{2M}
\big[\frac{\partial^2}{\partial R^2}+\frac{2}{R}\frac{\partial}{\partial R}
\big] +V^{(0)}(R)+\hat{H_0}(\xi)+V_c(R,\xi)
\end{equation}
where $V^{(0)}(R)$ is the bare interaction between the target and
projectile nuclei, ${\hat H}_0$ is the unperturbed Hamiltonian of the
electrons, and $V_c(R,\xi)$ is the interaction between the electrons
and nuclei. 
Denoting the wave function of electrons and the distance between the
projectile and the target at time $t$ by $\phi(\xi,t)$ and $R(t)$,
respectively, the time dependent Schr\"odinger equation and the classical
Newtonian equation for them read 
\begin{eqnarray}
i\hbar \frac{\partial \phi(\xi,t)}{\partial t} &=& \left[ H_0(\xi)+V_c(R(t),
\xi)\right] \phi(\xi,t)  \label{eq:clarel} \\
M\frac{d^2 R(t)}{dt^2} &=& -\frac{d}{dR} \left[
V^{(0)}(R)+\Delta V(R) \right]  \label{eq:clarnu}
\end{eqnarray}
where 
\begin{eqnarray}
\Delta V(R)=\langle \phi | \left[ H_0(\xi) + V_c(R(t),\xi)\right] | \phi
\rangle  \label{scpot}
\end{eqnarray}
Eqs.(\ref{eq:clarel}) and (\ref{eq:clarnu}) lead to the following energy
conservation law. 
\begin{equation}
\frac{M}{2}(\frac{dR(t)}{dt})^2 +V^{(0)}(R(t)) +\Delta V(R)
=E  \label{eq:clare}
\end{equation}
Eqs. (\ref{eq:clarel}) through (\ref{scpot}) have been derived from the time
dependent Schr\"odinger equation for the total wave function $\Psi(R,\xi,t)$
by approximating $\Psi(R,\xi,t)$ by a product of the wave functions for the
relative motion between nuclei $\frac{1}{R}\chi(R,t)$ and that for electrons 
$\phi(\xi,t)$, and determining them by a variational principle, thus the name
of a mean field theory, and then converting the Schr\"odinger equation for 
$\chi(R,t)$ into Newtonian equation for $R(t)$ following the idea of
Ehrenfest, i.e. by assuming that the classical time dependent variable $R(t)$
and its conjugate momentum $P(t)$ are given by the expectation values of the
corresponding operators in the state $\chi(t)$ (see ref.\cite{tkb} for
details). We call $\Delta V(R)$ the screening potential, which is nothing
but the total energy of the electrons at each time $t$ or equivalently at
each $R(t)$. We note that eq.(\ref{eq:clarnu}) is equivalent to 
\begin{eqnarray}
M\frac{d^2 R(t)}{dt^2} &=& -\frac{d}{dR} V^{(0)}(R)
- \langle \phi | \left[\frac{\partial }{\partial R(t)} V_c(R(t),\xi)\right] 
| \phi\rangle  \label{eq:clarnub}
\end{eqnarray}
which is more familiar in literatures. We prefer to the expression 
in eq.(\ref{eq:clarnu}), because it connects more directly to the 
corresponding equation in the tunneling region which we show below. 

We determine the time evolution in the classically allowed region by solving
eqs.(\ref{eq:clarel}) and (\ref{eq:clarnu}) along the real time axis with
the proper initial condition. Once the velocity of the relative motion
becomes zero, we switch to the imaginary time, $t=-i\tau $, and continue to
follow the time evolution in the tunneling region using the following
equations, 
\begin{eqnarray}
\hbar \frac{\partial \phi (\xi ,\tau )}{\partial \tau } &=&-\left[ H_{0}(\xi
)+V_{c}(R(\tau ),\xi )\right] \phi (\xi ,\tau )  \label{eq:clfbel} \\
\hspace*{-3mm}M\frac{\partial ^{2}R(\tau )}{\partial \tau ^{2}} &=&\frac{
\partial }{\partial R}\left[ V^{(0)}(R)+\Delta V(R)\right] 
\label{eq:clfbrel}
\end{eqnarray}
The screening potential and the energy conservation law in the tunneling
region are given by, 
\begin{equation}
\Delta V(R)=\frac{\langle \phi |\left[ H_{0}(\xi )+V_{c}(R(\tau ),\xi )
\right] |\phi \rangle }{\langle \phi |\phi \rangle }.  \label{eq:scpottu}
\end{equation}
\begin{equation}
-\frac{M}{2}(\frac{\partial R(\tau )}{\partial \tau })^{2}+V^{(0)}(R(\tau
))+\Delta V(R)=E  \label{eq:clfbe}
\end{equation}
We note that the norm of the wave function of electrons is not conserved in
the tunneling region. Accordingly, the denominator of the screening
potential given by the r.h.s. of eq.(\ref{eq:scpottu}) 
is essential as we see later.
 We note also that the potential renormalization 
given by eq.(\ref{eq:scpottu}) is the equivalent potential of the dynamical 
norm factor, which has been introduced in ref.\cite{dn} in order to 
take non-adiabatic effects into account to correct the calculation 
of the tunneling probability in the adiabatic approximation 
(see ref.\cite{tkb} for the derivation of these equations and the details.).   

Using the screening potential in the tunneling region thus obtained, we
calculate the tunneling probability in the presence of electrons by the
following WKB formula 
\begin{eqnarray}
P(E) &=&\exp \left( -2\sqrt{\frac{2M}{\hbar ^{2}}}\int_{R_{b}}^{R_{a}}dR%
\sqrt{V^{(0)}(R)+\Delta V(R)-E}\right)   \nonumber \\
&=&\exp \left( \frac{-4}{\hbar }\int_{\tau _{a}}^{\tau _{b}}d\tau \lbrack
V^{(0)}(R)-E]\right) \exp \left( \frac{-4}{\hbar }\int_{\tau _{a}}^{\tau
_{b}}d\tau \Delta V(R)\right)   \label{eq:ppre}
\end{eqnarray}
where $R_{a}$ and $R_{b}$ are the classical turning points on both sides of
the effective potential barrier $V^{(0)}(R)+\Delta V(R)$, and $\tau _{a}$
and $\tau _{b}$ are the corresponding times along the imaginary time axis. We
then convert the enhancement factor $f=\frac{P(E)}{P_{0}(E^{\prime })}$,
where $P_{0}(E^{\prime })$ is the tunneling probability in the absence of
electrons, into a screening energy using the relation 
\begin{eqnarray}
U_{e}=\frac{E_{K}^{\infty }}{\pi \eta (E_{K}^{\infty })}\log {\left( \frac{
P(E)}{P_{0}(E^{\prime })}\right) }=\frac{E_{K}^{\infty }}{\pi \eta
(E_{K}^{\infty })}\log {\left( \frac{P(E_{K}^{\infty }+\epsilon ^{(i)})}{
P_{0}(E_{K}^{\infty })}\right) }
\label{eq:scenergy}
\end{eqnarray}
where $\eta (E)$ is the Sommerfeld parameter, 
$E_{K}^{\infty }$ is  the kinetic energy of the relative motion between
the target and projectile nuclei and  $\epsilon ^{(i)}$  is the total energy
of electrons in the center of mass system in the initial asymptotic region.  
The latter is identical with the screening potential $\Delta $V at the 
initial time and is given by 
\begin{equation}
\epsilon ^{(i)}=\frac{1}{2}\mu _{e}v_{T}^{2}+\epsilon _{T}
\label{eq:initeel}
\end{equation}
where $\epsilon _{T}$ is the binding energy of electrons in the initial
state in the target atom and $v_{T}=\frac{M_{P}}{M_{T}+M_{P}}v_{\infty }$,
$v_\infty$ being $\frac{\partial R(t)}{\partial t}$ at $t=-\infty$, 
is the velocity of the target nucleus 
relative to the center-of-mass of the projectile and target nuclei 
at the initial time. The reduced mass $\mu_e$ is given by 
$\frac{1}{\mu _{e}}=\frac{1}{m_{e}}+\frac{1}{M_{T}+M_{P}}$. 
Strictly speaking $\epsilon_T$ on the r.h.s. of eq.(\ref{eq:initeel}) 
should be replaced by 
$\epsilon'_T=\epsilon_T \times \frac{M_P+M_T}{m_e+M_P+M_T}\times 
\frac{m_e+M_T}{M_T}$. However, 
the difference between $\epsilon'_T$ and $\epsilon_T$ can be ignored in the 
present problem. 
Note that we compare the
tunneling probabilities for the same kinetic energy of relative motion of
the nuclei in the presence and in the absence of the electrons.
That is why we use different notations for the energy arguments 
in the barrier penetrability in the second term of eq.(\ref{eq:scenergy}).
Also, we use in eq.(\ref{eq:scenergy}) and in what follows the lower index $0$ 
to denote the barrier penetrability and the cross section calculated 
in a two body system and distinguish them from the corresponding quantities 
calculated including electrons. 

We now apply our formalism to D+d and $^{3}$He$^{+}$+d reactions. We choose
these systems for simplicity of the treatment because  the screening effects
are due to a single electron. Moreover, there exists experimental data for
the D+d reaction \cite{ggjrz} at a low energy, i.e. at $E_{c.m.}$= 1.62 keV, 
though experiments have been performed for a molecular target 
rather than an atomic target \cite{ggjrz}.

\begin{figure}[hbt]
 \begin{center}
   \epsfig{file=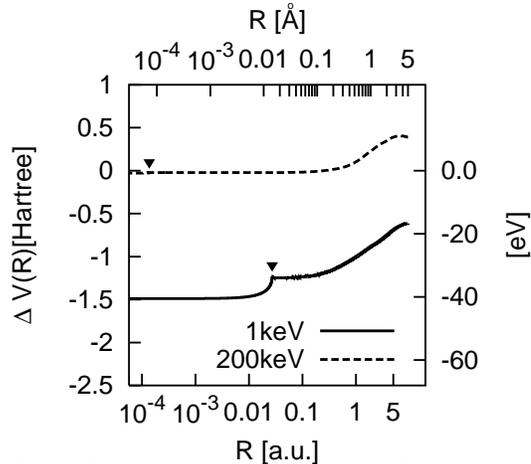,scale=0.7}
  \caption{Screening potential for the D+d reaction 
      at the center-of-mass energies 1 and 200 keV as a function of the 
      separation distance between the nuclei.  The filled triangles 
      show the position of the external classical turning point.}
  \label{fig:particle}
  \end{center}
\end{figure}    

Fig.1 shows the screening potential for the D+d reaction at $E_{c.m.}$=1 keV
(solid line) and 200 keV (dashed line). The asymptotic values and their
incident energy dependence can be understood 
from eq.(\ref{eq:initeel}). The closed
triangles show the external classical turning points. Two
interesting things can be noticed. The first is that the value of the screening
potential at the external turning point for 1 keV is -34.0 eV, which matches
with the average of 
binding energies $\epsilon_{UA}^{(g)}$=-54.4 eV 
in the lowest gerade and 
$\epsilon_{UA}^{(u)}$=-13.6 eV 
in the ungerade states, i.e. in the
1s- and 2p- states, of the united atom $^4$He$^+$. 
This indicates that the reaction takes place almost adiabatically 
in both gerade and ungerade configurations at this energy. The
second observation is that the screening potential for $E_{c.m.}=$ 1 keV 
changes very
fast just inside the external classical turning point. This can be
understood from eqs.(\ref{eq:scpottu}) and (\ref{eq:clfbel}) as a
consequence that the contribution to the mean potential from the ungerade
configuration, which has higher electronic energy, quickly dies out  as the
relative motion between the projectile and target penetrates into the
tunneling region. 

\begin{figure}[htbp]
  \begin{center}
    \epsfig{file=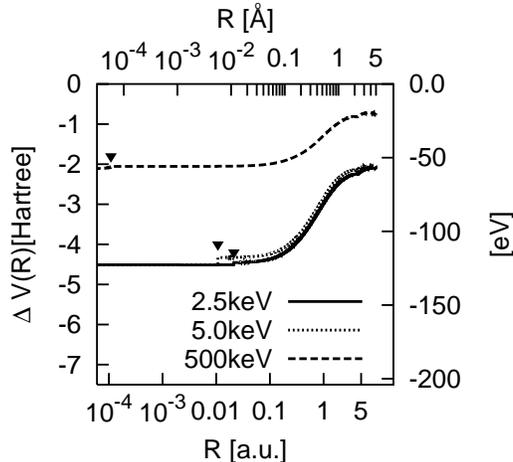,scale=0.7}
    \caption{The same as Fig.1, but for the $^3$He$^+$+d reaction 
      at three incident energies E$_{c.m.}$= 2.5, 5.0 and 500 keV.}
    \label{fig:Eel3Hed}
  \end{center}
\end{figure}

Fig.2 shows the screening potential for the $^{3}$He$^{+}$
+d reaction. Though it is not so drastic as that for the case of D+d
reaction, we see a similar structure near the external classical turning
point for the cases of 
$E_{c.m.}$=2.5(the solid line) and 5.0 keV(the dotted line). 
The screening potentials for these two energies 
merge for distances smaller than the external classical 
turning point for $E_{c.m.}$=5.0 keV.
In this system, the radial variation of the screening potential is caused 
by the admixture of excited states of electron 
due to pre-tunneling transitions, i.e. due to 
the transitions of the electronic state 
induced on the way from the initial asymptotic region 
to the classical turning point. 
One will then expect that the variation of the screening potential 
just after the system entered the tunneling region 
will get less significant at 
lower incident energies, because the process will become more and more 
adiabatic and the pre-tunneling electronic transition 
will become less important. 
This accords with Fig.2, where one sees that 
the variation of the screening potential at $E_{c.m.}$=2.5 keV is much 
smaller than that for $E_{c.m.}$=5.0 keV. This contrasts to the case for 
D+d reaction, where the symmetry of 
the system admixes the gerade and ungerade states with equal weight 
at any incident energies including the low energy adiabatic limit.
Note that eqs.(\ref{eq:clfbel}) and (\ref{eq:scpottu}) 
indicate that the screening potential at low energies 
will converge to the value in the adiabatic limit 
given by the binding energy of the 1s state of 
electron in the united atom 
as the tunneling process proceeds, and thus explains why 
the screening potentials 
for $E_{c.m.}$=2.5(the solid line) and 5.0 keV(the dotted line) merge 
for distances smaller than the classical turning point 
for $E_{c.m.}$=5.0 keV.
Note that the screening potential 
at $E_{c.m.}$=500 keV (the dashed line) 
shows no corresponding sharp radial variation and looks quite different 
from that for $E_{c.m.}$=2.5(the solid line) and 5.0 keV (the dotted line). 
This will be partly because the tunneling process at $E_{c.m.}$=500 keV 
is near to the sudden limit rather than to the adiabatic limit, 
and also because the tunneling region is much smaller. 
Note also that the abscissa is log-scale. It makes the variation at  
larger values of R appear more drastic.
The situation is similar in the case of 
D+d fusion reaction at $E_{c.m.}$=200 keV shown in Fig.1. 

\begin{figure}[htbp]
  \begin{center}
    \epsfig{file=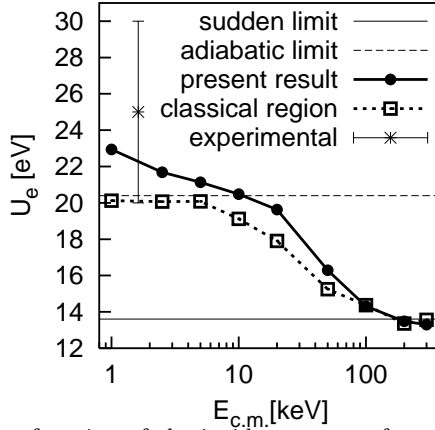,scale=0.7}
    \caption{Screening energy $U_e$ 
      as a function of the incident center-of-mass energy for 
      the D+d reaction. The experimental value is for a molecular deuteron 
      target taken from ref.14.}
    \label{fig:Ddscrnen}
  \end{center}
\end{figure}

Fig.3 shows the screening energy for the D+d reaction. The closed circles
are the results of our method. The open squares have been calculated in the
same way as in ref.\cite{ShoppaKooninLangankeSeki}. The horizontal solid and
dashed lines are the screening energies in the sudden and adiabatic reaction 
limits, $U_{e}^{(S)}$ and $U_{e}^{(AD)}$, respectively, which are given by 
\begin{eqnarray}
U_{e}^{(S)} &=&\frac{M_{T}}{M_{P}+M_{T}}\times 2 \times Z_{P}Z_{T} 
\times \epsilon _{H},
\label{uef1} \\
&=&\frac{M_{T}}{M_{P}+M_{T}}\times 2 \times Z_{P}Z_{T} \times 13.6eV=13.6eV.
\label{uef2} \\
U_{e}^{(AD)} &=& \epsilon_T-\epsilon_{UA}
\label{ueada}\\
&=&\frac{1}{2}\left[ \left( 54.4-13.6\right) +\left(13.6-13.6\right)\right] eV
\label{ueadb} \\
&=&20.4eV
\end{eqnarray}
These formulas, eqs.(\ref{uef1}) and (\ref{ueada}), for limiting cases 
are derived under the assumption that the tunneling region is much 
smaller than the Bohr radius of the united atom. Moreover, eq.(\ref{uef1}) has 
assumed that the screening electron occupies the 1s state of the target atom
(see ref.\cite{tkb} for details). 
In eq.(\ref{uef1}), $\epsilon _{H}=13.6$ eV is the binding energy 
of the 1s orbit in the Hydrogen atom. In eq.(\ref{ueada}), $\epsilon_T$ 
and $\epsilon_{UA}$ are the binding energies of the electron in the 
target and united atoms, respectively. The electron is assumed to 
occupy the adiabatic state with the same label $i$ in both atoms because 
of the slow adiabatic process. In the second line of $U_{e}^{(AD)}$, 
i.e. in eq.(\ref{ueadb}), we have used the actual values in the 
present case by taking the symmetry property of the D+d system into account. 
 As one expects, the screening energy
converges to that in the sudden reaction limit at high energies. It converges
to the adiabatic limit at low energies if one calculates in the way of ref.
\cite{ShoppaKooninLangankeSeki} by studying only the classical region. 
The star with error bar is the experimental value taken from ref.\cite{ggjrz}.
However, this should be taken merely as a reference, because as mentioned 
before the experiments have been performed not for an atomic deuteron target,
but for a molecular deuteron target. 
In this connection, we wish to mention that 
ref.\cite{shoppa96} has shown that the screening effect 
for the molecular D$_2$ target is larger than that for the 
atomic D target and attributed this difference to 
the reflection symmetry of the d+D system.  

The remarkable thing is that our calculations give systematically  
a larger screening energy than that in the conventional calculations. 
 At low energies, this can be understood  
in the following way. Using the screening potential at the external 
turning point $R_t$, 
the enhancement factor is calculated in the conventional method,
e.g. in ref.\cite{ShoppaKooninLangankeSeki}, by 
\begin{eqnarray}
f_{c} &=&\frac{\sigma_0 (E_{K}^{\infty }+\epsilon^{(i)}- 
\Delta V(R_{t}))}{\sigma
_{0}(E_{K}^{\infty })}  \label{fca} \\
&\approx &\frac{\sigma_0 (E_{K}^{\infty }+\epsilon_T- 
\Delta V(R_{t}))}{\sigma
_{0}(E_{K}^{\infty })}  \label{fcb} \\
&\approx &\frac{\sigma_0 (E_{K}^{\infty }+(U_{e}^{(g)}+U_{e}^{(u)})/2)}
{\sigma_{0}(E_{K}^{\infty })}.  \label{fcc} 
\end{eqnarray}
In transforming from eq.(\ref{fca}) to eq.(\ref{fcb}), 
we have ignored the difference between $\epsilon^{(i)}$ and $\epsilon_T$ 
given by eq.(\ref{eq:initeel}) in accord with the adiabatic process.
Also, in order to move further to eq.(\ref{fcc}),
we have used the fact, 
which we remarked before concerning Fig.1, that the screening potential 
at the external classical turning point can be understood 
in terms of the binding energies of the electron in the gerade 
and ungerade configurations of the united atom. 
On the other hand, our method, which handles the tunneling region explicitly,
leads to 
\begin{eqnarray}
f_{t} &=&\frac{\sigma ^{(g)}+\sigma ^{(u)}}{2\sigma _{0}} \\
&=&\frac{\sigma_0 (E_{K}^{\infty }+U_{e}^{(g)})+\sigma_0 (E_{K}^{\infty
}+U_{e}^{(u)})}{2\sigma _{0}(E_{K}^{\infty })}  \label{eq:enhwmf}
\end{eqnarray}
for the enhancement factor. These equations can be derived from 
eqs.(\ref{eq:clfbel}),(\ref{eq:scpottu}) and (\ref{eq:ppre}) 
by assuming that there are no change with the adiabatic energies and the
adiabatic states in the tunneling region. This assumption will be reasonable
in the present case, for which the tunneling region is much smaller than the
Bohr radius of the united atom as shown in Table 1. 
Since the excitation function of the fusion 
cross section is a convex increasing function of 
the incident energy, $f_t$ is larger 
than $f_c$. The conventional method thus underestimates the screening 
energy. 

Fig.3 clearly exemplifies this effect. It is important to
properly calculate the enhancement factor in order to get a reliable value
of the screening energy. This can be achieved either by explicitly handling
the tunneling region like in our method, or by studying the distribution of
the electronic state over different adiabatic states at the external 
classical turning point, and calculate the fusion probability for each of 
them and taking average afterwards with the proper weight. 
In this respect, we note that our numerical result at $E_{c.m.}$=1.6 keV 
agrees with the value given by eq.(\ref{eq:enhwmf}).
Assume that the electronic wave
function is distributed over different adiabatic states $\varphi _{n}$,
whose corresponding binding energy $\epsilon_T-U_n$, with
probability $P_{n}$, then eqs.(\ref{eq:clfbel}), (\ref{eq:scpottu}) and 
(\ref{eq:ppre}) lead in general to
\begin{eqnarray}
f_{t}=\frac{\Sigma _{n}P_{n}\times \sigma_0 (E_{K}^{\infty }+U_{n})}{\sigma
_{0}(E_{K}^{\infty })}
\label{ftstg}
\end{eqnarray}
which should be compared with the formula 
\begin{eqnarray}
f_{c}=\frac{\sigma_0 (E_{K}^{\infty }+\Sigma _{n}P_n\times U_{n})}{\sigma
_{0}(E_{K}^{\infty })}
\label{fcstg}
\end{eqnarray}
in the conventional method, where one first calculates the average 
potential and then calculates the tunneling probability for it. 
We have assumed that the tunneling region is
much smaller than the Bohr radius of the united atom and ignored the change
of the adiabatic states and their corresponding energies in the tunneling
region. Note that the presence of the denominator on the r.h.s. of 
eq.(\ref{eq:scpottu}) is essential in deriving eq.(\ref{ftstg}).

\begin{table}[htbp]
    \caption{\label{EofRDd}Incident energy dependence of the external classical turning 
    point for the D+d reaction. The initial velocity of the relative motion 
    in the unit of the 
      light speed $c$, $v_{\infty}/c$, the 
      external turning point in the absence of an electron $R_t^{(0)}$ and
      in its presence $R_t$, 
      and the mean square radius of the electron 
      at the external classical turning point are shown 
      for three bombarding energies.}
    \catcode`?=\active \def?{\phantom{0}}
    \renewcommand{\arraystretch}{1.2}
    \begin{tabular}{c|l|l|l|c}                
       $E_K^{\infty}$(keV) & $v_{\infty}/c$ & $R_t^{(0)}$(\AA) & $R_t$(\AA) & Electron Radius(\AA)\\ \hline
       100.0          & 0.0146?        & 0.000144?          & 0.000144?  & 0.456   \\  [0pt]
       ?10.0          & 0.00462        & 0.00144??          & 0.00144??  & 0.518   \\  [0pt]
       ??1.0          & 0.00146        & 0.0144???          & 0.0142???  & 0.688   
    \end{tabular}
\end{table}

In summary, we have presented a semiclassical mean field theory 
of quantum tunneling which treats both classical and tunneling regions 
in a consistent way. Applying the formalism to the problem of screening 
effects by bound target electrons in low energy nuclear reactions in 
laboratories, we have shown that 
the screening potential shows a characteristic radial variation contrary to 
the assumption of a constant potential shift in all previous analyses. 
We have shown also that the proper treatment of the tunneling region 
leads to an increase of the 
screening energy compared with that estimated in the previous 
mean field theory, 
which studies only the classical region and calculates the tunneling 
probability by using the average potential 
at the external classical turning point.  
The above effects are, however, too small 
to explain the large experimental screening energies reported 
in ref.\cite{rolfs95}. Remember in this connection that 
the large screening energy 
obtained in this study for the 
D+d reaction has been caused by the symmetry special to this system, 
and cannot be generalized to other systems.

The authors thank Bertrand Giraud for useful discussions.
This work has been partly performed while 
the authors stayed at the INFN at Catania. 
We thank M. Di Toro, A. Bonasera, C. Spitaleri and S. Terranova 
for useful discussions and warm hospitality. 
This work is supported by
the Grant-in-Aid for Scientific Research from 
Ministry of Education, Culture, Sports, Science and Technology 
under Grant No. 12047203 and No. 13640253 
and by the Japan Society for the Promotion 
of Science for Young Scientists under the contract No. 12006231.


\end{document}